\RequirePackage[2020-02-02]{latexrelease}
\documentclass[aps,preprint,11pt]{revtex4}
\usepackage{amsmath}
\usepackage{graphicx}
\usepackage{amsfonts}
\usepackage{amssymb}
\usepackage{blindtext}
\usepackage{multirow}
\usepackage{booktabs}
\usepackage[T1]{fontenc}
\usepackage[utf8]{inputenc}
\usepackage{float}
\usepackage{natbib}
\usepackage{notoccite}
\DeclareMathOperator{\arctanh}{arctanh}

\DeclareMathOperator{\Li}{Li}

\RequirePackage[2020-02-02]{latexrelease}
\RequirePackage[2020-02-02]{latexrelease}
\providecommand{\U}[1]{\protect\rule{.1in}{.1in}}
\providecommand{\U}[1]{\protect\rule{.1in}{.1in}}
\providecommand{\U}[1]{\protect\rule{.1in}{.1in}}
\providecommand{\U}[1]{\protect\rule{.1in}{.1in}}
\providecommand{\U}[1]{\protect\rule{.1in}{.1in}}

\makeatletter
\renewcommand{\citation}[1]{%
	\g@addto@macro{\citation@list}{,#1}%
}
\newcommand*{\citation@list}{} % initialize
\newcommand{\sortbibitem}[2]{%
	\global\@namedef{bibitem@#1}{%
		\bibitem{#1} #2
	}%
}
\newcommand{\sort@bibitems}{%
	\@for\next:=\citation@list\do{%
		\@nameuse{bibitem@\next}%
		\global\@namedef{bibitem@\next}{}%
	}%
}
\expandafter\def\expandafter\endthebibliography\expandafter{%
	\expandafter\sort@bibitems\endthebibliography
}
\makeatother
	
\begin{document}

\title{Thermodynamic properties of an ideal Quark-Gluon plasma under quantum gravitational effects }

\author{Djamel Eddine Zenkhri $^{\text{a}}$}
	\email{zenkhridjameleddine@gmail.com,dzenkhri@univ-ouargla.dz}
\author{Abdelhakim Benkrane$^{\text{a}}$}
        \email{abdelhakim.benkrane@univ-ouargla.dz }
 \affiliation{$^\text{a}$Laboratoire LRPPS, Faculté des Mathématiques et des Sciences de la Matière.  Université Kasdi Merbah Ouargla, Ouargla 30000, Algeria
 }

\begin{abstract}
In this study, we investigate the thermodynamic properties of an ideal Quark-Gluon Plasma (QGP) at a vanishing chemical potential, under the influence of quantum gravitational effects, specifically incorporating the Linear-Quadratic Generalized Uncertainty Principle (LQGUP). We analyze the impact of LQGUP on key thermodynamic quantities, including the grand canonical potential, pressure, energy density, entropy, speed of sound, and the bulk viscosity's response to changes in the speed of sound. %Additionally, we examine the time evolution of the universe's temperature in this context.
Furthermore, we extend our analysis to examine the time evolution of the universe's temperature in the presence of LQGUP effects. 
%\textst{Our findings provide new insights into the behavior of QGP under extreme conditions and contribute to a deeper understanding of the interplay between quantum gravity and high-energy plasma physics.} 

     \textbf{keywords:} Quark-Gluon plasma, Generalized uncertainty principle, Thermodynamics properties.
		\end{abstract}
			\maketitle

		\section{Introduction}
A deeper understanding of the universe depends on uncovering matter's fundamental forces and behavior under extreme conditions. According to Friedmann's solution \cite{friedman1922krummung} of Einstein's gravitational equation, the universe experienced an expansion from a singularity point at time zero which has been confirmed by the formulation of Hubbel's law for the redshift of distant galaxies \cite{hubble1929relation}. 
At the dawn of the universe, just microseconds after the Big Bang, a state of matter unlike anything observed today is thought to have existed: quark-gluon plasma (QGP). This phase consists of unconfined quarks and gluons, the elementary building blocks of protons and neutrons, which exist freely rather than confined within individual particles.  It can be recreated in high-energy nuclear collisions, such as those conducted at the Relativistic Heavy Ion Collider (RHIC) and the Large Hadron Collider (LHC). Recent experimental results suggest that these collisions have successfully produced a QGP with near-perfect fluid properties, making studying its thermodynamic and transport characteristics increasingly pertinent. Throughout the universe's evolution, the quark-gluon plasma (QGP) transitions into hadronic matter, a process governed by quantum chromodynamics (QCD), which describes the strong interactions \cite{yagi2005quark}. The key distinction between the hadronic and QGP phases lies in the relative significance of short-range and long-range interactions among their constituents across the expected phase transition. In the hadronic phase, short-range interactions among hadrons dominate, which can be described by Boltzmann-Gibbs statistics. In contrast, the QGP phase is marked by a significant reduction in short-range interactions due to "asymptotic freedom," leading to a predominance of long-range interactions \cite{teweldeberhan2003generalized}.

%The Generalized Uncertainty Principle (GUP), is one of the most exciting predictions of some approaches related to quantum gravity
In recent years, the Generalized Uncertainty Principle (GUP) has garnered significant attention as an extension of the traditional Heisenberg Uncertainty Principle (HUP), motivated by the need to reconcile quantum mechanics with gravitational effects \cite{snyder1947,yang1947,mead1964,karolyhazy1966}. The foundational theories, such as string theory, loop quantum gravity, deformed special relativity, and black hole physics, have all contributed to the development of various forms of GUP, each characterized by a parameter $\beta$, which can be derived either theoretically \cite{amati1987,gross1987,amati1989,konishi1990} or phenomenologically \cite{das2008,das2010,das2009,farag2012}. Kempf's works are considered pioneering in the GUP \cite{kempf1995hilbert, kempf1994uncertainty, hinrichsen1996maximal}, and one of his main motivations was the necessity of a minimal length within the framework of quantum gravity and string theory.

The GUP model predicts a maximum observable momentum and a minimal measurable length. Accordingly, and via the Jacobi identity $\left[x_i,x_j\right]=\left[p_i,p_j\right]=0$  results in \cite{ali2009}:
\begin{equation}
\left[x_i,p_j\right]=i\hbar\left[\delta_{ij}-\beta\left(p\delta_{ij}+\frac{p_ip_j}{p}\right)+\beta^2(p^2\delta_{ij}+3p_ip_j)\right]
\label{dif}
\end{equation}
where $\beta=\alpha_0/M_pc=\alpha_0l_p/\hbar$. $M_p=4.34\times 10^{-9}\text{kg}$, $l_{p}=10^{-35}\text{m}$ and $c=3 \times 10^{8} \text{m}/\text{s}$ are Planck mass and length and speed of light, respectively. $\alpha_0$ sets on the upper and lower bounds to $\alpha$. 
Eq.(\ref{dif}) leads to a minimal measurable length $\Delta x_{min}\approx \alpha_0 l_p$ suggesting that spacetime has a discrete nature \cite{ali2009}, and  maximum measurable momentum $\Delta p_{max}\approx \frac{M_p c}{\alpha_0}$. Given the fact that the GUP leads to a minimum length, it is effective at microscopic scales and very high energies. As a significant consequence, we can infer that the Generalized Uncertainty Principle plays a very important role in the early stages of the universe \cite{adler1999gravity, zhu2009influence, hassanabadi2020investigation}.  Recently, it was proved that the GUP may naturally arise from the localized quantum mechanical wave function, using a horizon wave function of Gaussian form \cite{casadio2014horizon, casadio2013localised, casadio2015inner}. Considering all these considerations, one can assert that the GUP is not merely a purely mathematical model but has significant physical implications and meanings. In the context of GUP,  there are many applications, the impact of GUP on Friedman equations \cite{majumder2011generalized}, Kratzer molecular potential \cite{bouaziz2015kratzer},   black hole remnants \cite{adler2001generalized}, power law central potential and hydrogen atom \cite{perivolaropoulos2017cosmological, brau1999minimal}, Unruh effect and related Unruh temperature \cite{scardigli2018modified}, quantum correction to the Coulomb potential \cite{baradaran2024some}, Wigner function of a classical harmonic oscillator in the configuration space \cite{parida2024gup}, and one-and two-dimensional Dirac oscillator \cite{dagoudo2024algebraic}.

%One notable application of GUP is its impact on the thermodynamic properties of the Quark Gluon Plasma (QGP) \cite{el2013,elmashad2012,abou2015}. It can be recreated in high-energy nuclear collisions, such as those conducted at the Relativistic Heavy Ion Collider (RHIC) and the Large Hadron Collider (LHC). Recent experimental results suggest that these collisions have successfully produced a QGP with near-perfect fluid properties, making studying its thermodynamic and transport characteristics increasingly pertinent.

Our goal in this paper is to treat the impact of GUP with the form \cite{ali2009}:
\begin{equation}
\Delta x \Delta p\geq \dfrac{1}{2}\left(1+\beta \Delta p+ \beta^2 \Delta p ^2 \right),
\label{eqdif}
\end{equation}
the equation of state, energy density, entropy, sound speed, and bulk viscosity ratio was calculated on the QGP phase.  Previous studies have explored the influence of GUP on the equation of state of an ideal QGP, revealing modifications in its thermodynamic quantities \cite{el2013,abou2015,elmashad2012,demir2018effect, barman2023qgp}. 
Based on the information above, there are significant correlations between the GUP and QGP. For instance, both are crucial in the early stages of the universe. Additionally, the QGP represents the state of matter at extremely high temperatures,  Consequently, the QGP phase is anticipated to undergo corrections at the Planck scale
\cite{barman2023qgp}. Motivated by these correlations and previous works and as a continuation of this line of research, we aim in our article to investigate the Linear-Quadratic Generalized Uncertainty Principle (LQGUP)-induced corrections to the thermodynamic properties of an ideal QGP. We neglect the third order of deformation
parameter $\beta$. \par 
%\textcolor{red}{It is crucial to note that the GUP expression (\ref{dif}) and (\ref{eqdif}) are not unique, there are other forms, for example, the quadratic GUP \cite{kempf1995hilbert}, Linear GUP \cite{ali2024covariant}, and a higher order GUP  \cite{pedram2012higher}. Moreover, it is also important to point out that, since the effects of LQGUP are compatible with doubly special relativity (DSR) \cite{cortes2005quantum, ali2009, vagenas2019linear}, and this latter leads to corrections in tests of special relativity, with the possibility of future experiments translating these effects, or perhaps they have already been observed in the high-energy tail of the cosmic ray spectrum \cite{cortes2005quantum, amelino2005phenomenology}, this may lead to possible the experimental observable to the physical properties we will calculate, which adds significant value to our work, extending beyond the theoretical aspect, particularly in studying the impact of LQGUP on the thermodynamic properties of QGP, for instance.}

It is important to emphasize that the GUP expressions (\ref{dif}) and (\ref{eqdif})are not unique; other forms have been studied, such as the quadratic GUP \cite{kempf1995hilbert}, Linear GUP \cite{ali2024covariant}, and higher-order GUP  \cite{pedram2012higher}. Furthermore, the LQGUP framework is consistent with Doubly Special Relativity (DSR) \cite{cortes2005quantum, ali2009, vagenas2019linear}, which introduces corrections to tests of special relativity. These corrections may become experimentally observable, potentially reflected in phenomena such as the high-energy tail of the cosmic ray spectrum \cite{cortes2005quantum, amelino2005phenomenology}, This connection enhances the significance of our study, as it not only focuses on the theoretical implications of LQGUP but also opens avenues for exploring its impact on the thermodynamic properties of QGP in experimental contexts.

This paper is organized as follows: Section 2 focuses on calculating the partition function of the quark-gluon plasma (QGP) under the framework of the Linear Quadratic Generalized Uncertainty Principle (LQGUP), from which the grand canonical potential, pressure, and energy density are derived. In Section 3, we investigate the influence of LQGUP on thermodynamic properties such as entropy and the speed of sound. Section 4 examines the impact of quantum gravity corrections, through the Generalized Uncertainty Principle (GUP), on the bulk viscosity of QGP. Section 5 explores the cosmological implications of the QGP phase, particularly the evolution of energy density and temperature in the early universe. Finally, the conclusions are summarized in Section 6.

\section{Equation of state and energy density }
The QGP is a state of matter where quarks and gluons are no longer confined within hadrons but instead move freely in a hot and dense medium, where at high temperatures, the typical momenta of quarks and gluons are large, and the running coupling $\alpha_{s}$ becomes weak due to asymptotic freedom \cite{yagi2005quark}. Quarks are the fundamental building blocks of matter and belong to fermions (half-integer spin e.g., 1/2, 3/2), nevertheless, gluons have a spin of 1, which is a characteristic of all bosons. In this section, we derive the thermodynamic properties (pressure, energy density, entropy...) of QGP in the case of bosons and fermions considering the GUP form of Eq. (\ref{eqdif}).

\subsection{Case of Bosons.}
The grand-canonical partition function for non-interacting massive bosons with $g$ internal degrees of freedom, finite temperature $T$ and chemical potential $\mu$, is given by:

\begin{align}
Z_{B} &=\prod_{k} \left[ \sum_{l=0}^{\infty }\exp \left( -l\frac{E\left(
k\right) -\mu }{T}\right) \right] ^{g},  \\
&=\prod_{k} \left[ 1-\exp \left( -\frac{E\left( k\right) -\mu }{T}\right) %
\right] ^{-g}.\label{eq4}
\end{align}%
In this context, the infinite product is computed over all conceivable momentum states, where $l$ denotes the occupation number associated with each quantum state characterized by its energy $E(k)=\sqrt{k^2+m^2}$ with mass $m$ and $k$ is the
momentum of the particle. 

For a particle of mass $M$ having a distant origin and an energy comparable
to the Planck scale, the momentum would be the subject of a
tiny modification so that the dispersion relation would too.
%According to the GUP approach, the dispersion relation reads
The effect of the aforementioned GUP on the relativistic dispersion relation is given by \cite{majhi2013modified}:
\begin{equation}
    E^2(k)=k^2c^2(1-\beta k+\beta^2 k^2)^2+M^2c^4,
\end{equation}
where $M$ and $c$ are the mass of the particle and the speed of
light as introduced by Lorentz and implemented in special
relativity, respectively. 

To facilitate the analysis, we adopt the chiral limit ($M = 0$) and zero chemical potential $(\mu = 0)$. These assumptions are justified at high energies, as corroborated by experimental observations. Under these conditions, the dominant excitation within the hadronic phase becomes the massless pion, while the quark-gluon plasma (QGP) is characterized by the presence of massless quarks and gluons. We use natural units in which $\hbar=c=1$. Hence, we have:
\begin{equation}
E(k)=k(1-\beta k+\beta^2 k^2).
\end{equation}
Since $1-\beta k+\beta^2 k^2$ is positive for all values of $k>0$, there are no concerns about the possibility of the energy becoming negative.
For large volumes, the sum over all states of a single particle can
be rewritten in terms of an integral \cite{vagenas2019linear} (with $D=3$),
\begin{align}
\sum \rightarrow \frac{V}{\left( 2\pi \right) ^{3}}\int d^{3}k\rightarrow \frac{V}{2\pi ^{2}}\int_{0}^{\infty }\frac{k^{2}}{\left(
1-\beta k+\beta^2 k^2\right) ^{4}}dk. \label{ps}
\end{align}
The authors in Ref. \cite{vagenas2019linear}, prove that the phase-space (\ref{ps}) complies with Liouville's theorem.
Now, to calculate the thermodynamic properties of QGP, we first have to calculate the partition function. Which can be calculated for the system of an ideal QGP by working in the grand canonical ensemble. For gluons (bosons), and after including the Linear-quadratic GUP deformation, the grand canonical partition function can be written by: 
\begin{align}
\ln Z_{B} &=-\frac{Vg}{2\pi ^{2}}\int_{0}^{\infty }k^{2}\frac{\ln \left[
1-\exp \left( -\frac{E\left( k\right) }{T}\right) \right] }{\left( 1-\beta k+\beta^2 k^2\right) ^{4}}dk \nonumber \\
&=-\frac{Vg}{2\pi ^{2}}\int_{0}^{\infty }k^{2}\frac{\ln \left[ 1-\exp
\left( -k\left( 1-\beta k+\beta^2 k^2\right) /T\right) \right] }{\left( 1-\beta k+\beta^2 k^2\right)
^{4}}dk.
\label{eq8}
\end{align}

By expanding up to the second order of $\beta$, one can get:
\begin{align}
\ln(Z_{B})=-\frac{Vg}{2\pi^{2}}\int_{0}^{+\infty} k^{2}\ln\left(1-\exp\left(-k\frac{( 1-\beta k+\beta^2 k^2)}{T}\right)\right)\left( 1+4\beta
k+6\beta ^{2}k^{2}\right) dk. \label{eqq7}
\end{align}
Let $x=k\left(  1-\beta k+\beta^2 k^2\right) /T$, so that $dx=\left(1 -2\beta+3\beta^2 k\right)
dk/T$. The momentum $k$ as a function of $x$ variable can be approximated to the second order of $\beta $ as follows:%
\begin{eqnarray}
k &=&xT+\beta k^{2}-\beta ^{2}k^{3}, \nonumber\\
&=&xT+\beta x^{2}T^{2}\left( 1+\frac{\beta k^{2}}{xT}-\frac{\beta ^{2}k^{3}}{%
xT}\right) ^{2}-\beta ^{2}x^{3}T^{3}\left( 1+\frac{\beta k^{2}}{xT}-\frac{%
\beta ^{2}k^{3}}{xT}\right) ^{3}, \nonumber\\
&=&xT+\beta x^{2}T^{2}\left( 1+2\frac{\beta k^{2}}{xT}\right) -\beta
^{2}x^{3}T^{3}, \nonumber\\
&=&xT+\beta x^{2}T^{2}+2\beta ^{2}x^{3}T^{3}\left( 1+\frac{\beta k^{2}}{xT}-%
\frac{\beta ^{2}k^{3}}{xT}\right) ^{2}-\beta ^{2}x^{3}T^{3},\nonumber \\
 &=&xT+\beta x^{2}T^{2}+\beta ^{2}x^{3}T^{3}, \label{K}
\end{eqnarray}%
by substituting the expression (\ref{K}), and expanding $\frac{k^{2}\left( 1+4\beta k+6\beta ^{2}k^{2}\right) }{\left( 1-2\beta
k+3\beta ^{2}k^{2}\right) }$ up to the second order of $\beta$:
\begin{align}
\frac{k^{2}\left( 1+4\beta k+6\beta ^{2}k^{2}\right) }{\left( 1-2\beta
k+3\beta ^{2}k^{2}\right) }&=k^{2}+6\beta k^{3}+15\beta ^{2}k^{4}\nonumber\\
&=x^{2}T^{2}+8\beta x^{3}T^{3}+36\beta ^{2}x^{4}T^{4},
 \label{eqq11}
\end{align}%
one gets the following form of the partition function for bosons under the impact of LQGUP:  
\begin{eqnarray}
\ln Z_{B}&=&-\frac{VgT}{2\pi ^{2}}\int_{0}^{\infty} \ln \left( 1-\exp \left( -x\right) \right)
\left( x^{2}T^{2}+8\beta x^{3}T^{3}+36\beta ^{2}x^{4}T^{4}\right) dx \\
&=&\frac{\pi ^{2}}{90}VgT^{3}+24\frac{Vg}{\pi ^{2}}\beta T^{4}\zeta \left(
5\right) +\frac{32}{70}Vg\pi ^{4}\beta ^{2}T^{5}.
\end{eqnarray}

In the case of $\beta\rightarrow 0$, the above equation is reduced to the partition function for bosons without the LQGUP effect.

\subsection{Case of Fermions}
The grand canonical partition function for fermions is expressed by:
\begin{align}
Z_{F} &=\prod_{k} \left[ \sum_{l=0}^{\infty }\exp \left( -l\frac{E\left(
k\right) -\mu }{T}\right) \right] ^{g}  \\
&=\prod_{k} \left[ 1+\exp \left( -\frac{E\left( k\right) -\mu }{T}\right)
\right] ^{g}\label{eq14}.
\end{align}%
For fermions, the deformed partition function with LQGUP is: 

\begin{align}
\ln(Z_{F})=\frac{Vg}{2\pi^{2}}\int_{0}^{+\infty} \frac{k^{2}\ln\left(1+\exp\left(-k(1+\beta k)/T\right)\right)}{(1-\beta k+\beta^2 k^2)^4}dk,
\end{align}
and by expanding to the first order of $\beta$, one can get:
\begin{align}
\ln(Z_{F})=\frac{Vg}{2\pi^{2}}\int_{0}^{+\infty} k^{2}\ln\left(1+\exp\left(-k\frac{(1-\beta k+\beta^2 k^2)}{T}\right)\right)\left( 1+4\beta k+6\beta ^{2}k^{2}\right)dk. \label{eq16}
\end{align}

Following a procedure analogous to that used for bosons, one can derive the partition function for fermions in the context of the LQGUP to get:  %as the previous section and by taking $x=k\frac{(1-\beta k+\beta^2 k^2)}{T}$ and $dx=\frac{1}{T}(1-2\beta k+3\beta^2 k^2)dk$ in Eq (\ref{eq16}), we get: 
%\begin{equation}
%\ln(Z_{F})=\frac{VgT}{2\pi^{2}}\int_{0}^{+\infty}k^{2}\frac{\left( 1+4\beta k+6\beta ^{2}k^{2}\right) }{\left( 1-2\beta
%k+3\beta ^{2}k^{2}\right) }\ln\left[1+\exp\left(-x\right)\right]dx %\label{eq18}
%\end{equation}

%We have proved in Eq.(\ref{K}) that the momentum $k$ can be expressed as a function of $x$ in the first order of $\beta $ as follows:%
%\begin{equation}
%k \simeq xT+\beta x^{2}T^{2}+\beta ^{2}x^{3}T^{3} \label{eq 17}
%\end{equation}%
%and the first term of the integral equation is expressed by:
%\begin{align}
%\frac{k^{2}\left( 1+4\beta k+6\beta ^{2}k^{2}\right) }{\left( 1-2\beta
%k+3\beta ^{2}k^{2}\right) }=x^{2}T^{2}+8\beta x^{3}T^{3}+36\beta ^{2}x^{4}T^{4}
 %\label{eqq23}
%\end{align}%

\begin{eqnarray}
\ln Z_{F}&=&-\frac{VgT}{2\pi ^{2}}\int_{0}^{\infty} \ln \left( 1-\exp \left( -x\right) \right)
\left( x^{2}T^{2}+8\beta x^{3}T^{3}+36\beta ^{2}x^{4}T^{4}\right) dx, \\
&=&\frac{7}{8}\frac{\pi ^{2}}{90}VgT^{3}+\frac{45}{2\pi ^{2}}Vg\beta T^{4}\zeta \left(
5\right) +\frac{31}{70}Vg\pi ^{4}\beta ^{2}T^{5}.
\end{eqnarray}

In the case of $\beta\rightarrow 0$, the above equation is reduced to the partition function for bosons without the GUP effect.\par

%The total grand canonical partition function of the QGP state can be given by adding the grand partition functions coming from the contribution of bosons (gluons), fermions (quarks), and vacuum as follows:
In the context of QGP, the total grand canonical partition function is obtained by summing the individual grand partition functions of its constituents: bosons (Gluons) $\ln (Z_{G})$, fermions (Quarks) $\ln (Z_{Q})$, and the vacuum state $\ln (Z_{V})$, as follows:

\begin{equation}
    \ln Z_{QGP}=\ln Z_{B}+\ln Z_{F}+\ln Z_{V},
\end{equation}
where the vacuum partition function reads $\ln\left(Z_{V}\right)=-VB/T$, and $B$ is the bag  constant  in case of MIT Bag model \cite{johnson1975bag}. Once we find the partition function all the state equations of free massless quarks and gluons of the QGP  can be derived.

The grand potential is related to the partition function (we have set the chemical potential equal to zero, namely
$\mu = 0$) by: $\Omega=-T\ln{Z}$. So the grand potential for the QGP system can be written as: 
\begin{gather}
\Omega(T,V,0)=-T\left\{ \ln\left(Z_{f}\right)+\ln\left(Z_{B}\right)+\ln\left(Z_{v}\right)\right\}\nonumber\\
=-\left(g_g+\frac{7}{8}g_q\right)\frac{\pi^{2}V}{90} T^{4}-\beta\left(24g_g+\frac{45g_q}{2}\right)\frac{ VT^{5}\zeta \left(
5\right)}{\pi ^{2}}-\beta^{2}\bigg(32g_g+31g_q\bigg)\frac{V\pi ^{4}T^{6}}{70}-BV.
\label{eqq223}
\end{gather}

%\begin{figure}[h!]
%	\includegraphics[width=1\linewidth, height=0.48\textheight]{Omega.eps}\hfill
%	\caption{The variation of the grand potential as a function of temperature $T$}
%	\label{Omega}
%\end{figure}
 Since we got the grand potential for gluons and quarks, one can conclude the equation of state, where for grand canonical ensemble  $P=-\dfrac{\partial \Omega}{\partial V}$.
 In its simplest form, assuming non-interacting, massless constituents and vanishing conserved charges, the ordinary Bag Model Equation of State (BM EoS)  is expressed as:
$$ 
\epsilon (T)=\sigma_{SB}T^4+B, \quad \quad P(T)=\frac{\sigma_{SB}}{3}T^4-B,
$$
in our case, and by including the LQGUP corrections, based on Eq.(\ref{eqq223}), we get:
\begin{gather}
P %\frac{\pi ^{2}}{90}g_qT^{4}+24\frac{g_g}{\pi ^{2}}\beta T^{4}\zeta \left(
%5\right) +\frac{32}{70}g_g\pi ^{4}\beta ^{2}T^{4}+\frac{7\pi ^{2}}{720}g_qT^{5}+\frac{45g_q}{2\pi ^{2}}\beta T^{4}\zeta \left(
%5\right) +\frac{31}{70}g_q\pi ^{4}\beta ^{2}T^{5}-B\nonumber\\
=\left(g_g+\frac{7}{8}g_q\right)\frac{\pi^{2}}{90} T^{4}+\beta\left(24g_g+\frac{45g_q}{2}\right)\frac{ T^{5}\zeta \left(
5\right)}{\pi ^{2}}+\beta^{2}\bigg(32g_g+31g_q\bigg)\frac{\pi ^{4}T^{6}}{70}-B\nonumber\\
=\frac{\sigma_{SB}}{3}T^{4} +\beta\left(24g_g+\frac{45g_q}{2}\right)\frac{ T^{5}\zeta \left(
5\right)}{\pi ^{2}}+\beta^{2}\bigg(32g_g+31g_q\bigg)\frac{\pi ^{4}T^{6}}{70}-B, \label{pressure}
\end{gather}
where $\sigma_{SB}=\left(g_g+\frac{7}{8}g_q\right)\frac{\pi^{2}}{30}$ is the Stefan-Boltzmann (SB) constant. Based on Gibbs condition \cite{kapoyannis2007}, the critical point is obtained at the phase equilibrium, which is satisfied when $P_H(T_c)=P_{QGP}(T_c)$. Then, the bag pressure can be expressed as: 
\begin{gather}
B=\left(g_{QGP}-g_{\pi}\right)\frac{\pi^{2}}{90} T_{c}^{4}+\beta\left(24(g_{g}-g_\pi)+\frac{45g_q}{2}\right)\frac{ T_{c}^{5}\zeta \left(
5\right)}{\pi ^{2}}+\beta^{2}\bigg(32(g_g-g_\pi)+31g_q\bigg)\frac{\pi ^{4}T_{c}^{6}}{70}, 
\label{B}
\end{gather}
with $T_c$: refers to the critical temperature at which a phase transition occurs between the hadronic matter (where quarks and gluons are confined inside hadrons) and the deconfined Quark-Gluon Plasma phase (where quarks and gluons are free to move). Here, we took the chiral limit ($m_q=0$), which leads to the massless pions dominant excitations in the hadronic phase \cite{el2013}. $g_{QGP}$ in the Eq. (\ref{B}) is given as $g_{QGP}=g_{g}+\dfrac{7}{8}g_{q}.$

\begin{figure}[h!]
	\includegraphics[width=1\linewidth, height=0.48\textheight]{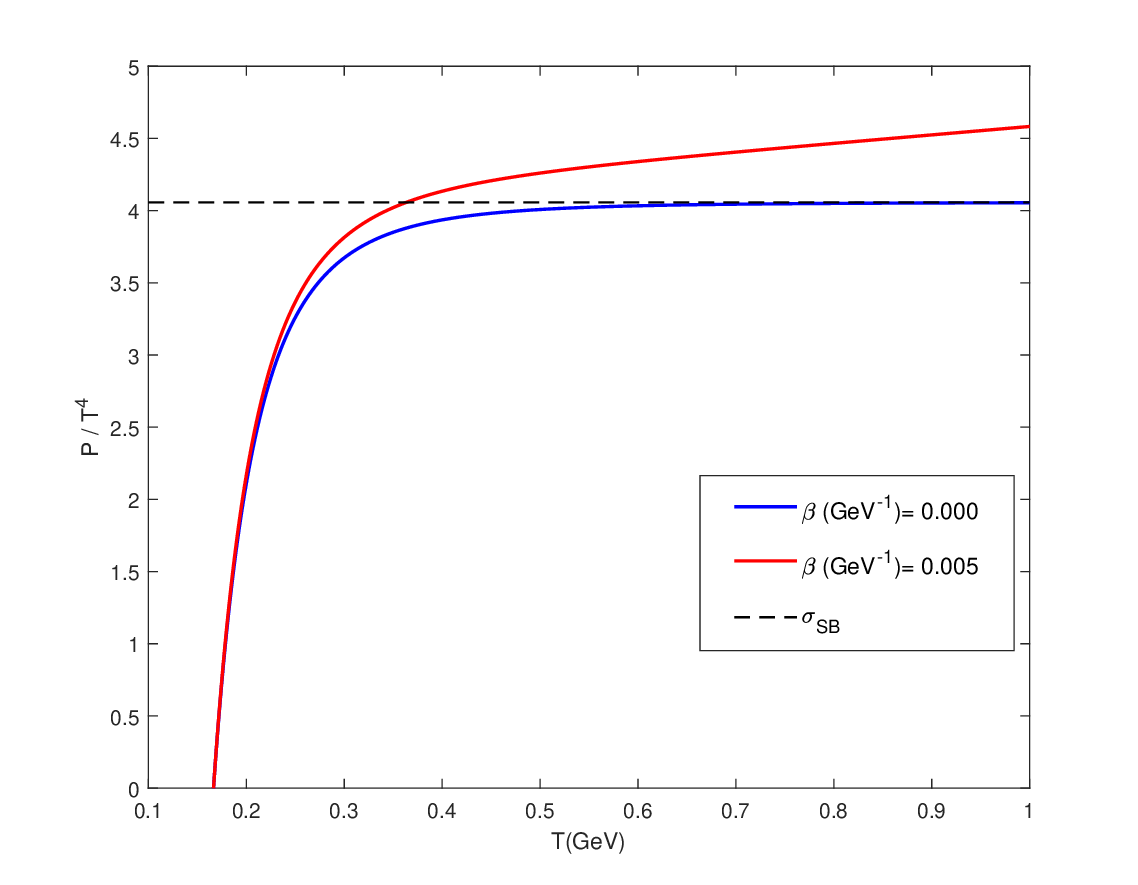}\hfill
	\caption{The pressure normalized to $T^4$ is given in dependence on T . The horizontal solid line represents the Stephan Boltzmann ($\sigma_{SB}$) limit. }
	\label{P}
\end{figure}

%We can explicitly add the result $\epsilon = 3P$ which is valid
%when the mass of particles is small relative to their energy (massless
%particles or ultra-relativistic gas).
As for density energy with GUP correction, we get:
\begin{gather}
\epsilon=\left(g_g+\frac{7}{8}g_q\right)\frac{\pi^{2}}{30} T^{4}+3\beta\left(24g_g+\frac{45g_q}{2}\right)\frac{ T^{5}\zeta \left(
5\right)}{\pi ^{2}}+3\beta^{2}\bigg(32g_g+31g_q\bigg)\frac{\pi ^{4}T^{6}}{70}+B. \label{density}
\end{gather}
\begin{figure}[h!]
	\includegraphics[width=1\linewidth, height=0.48\textheight]{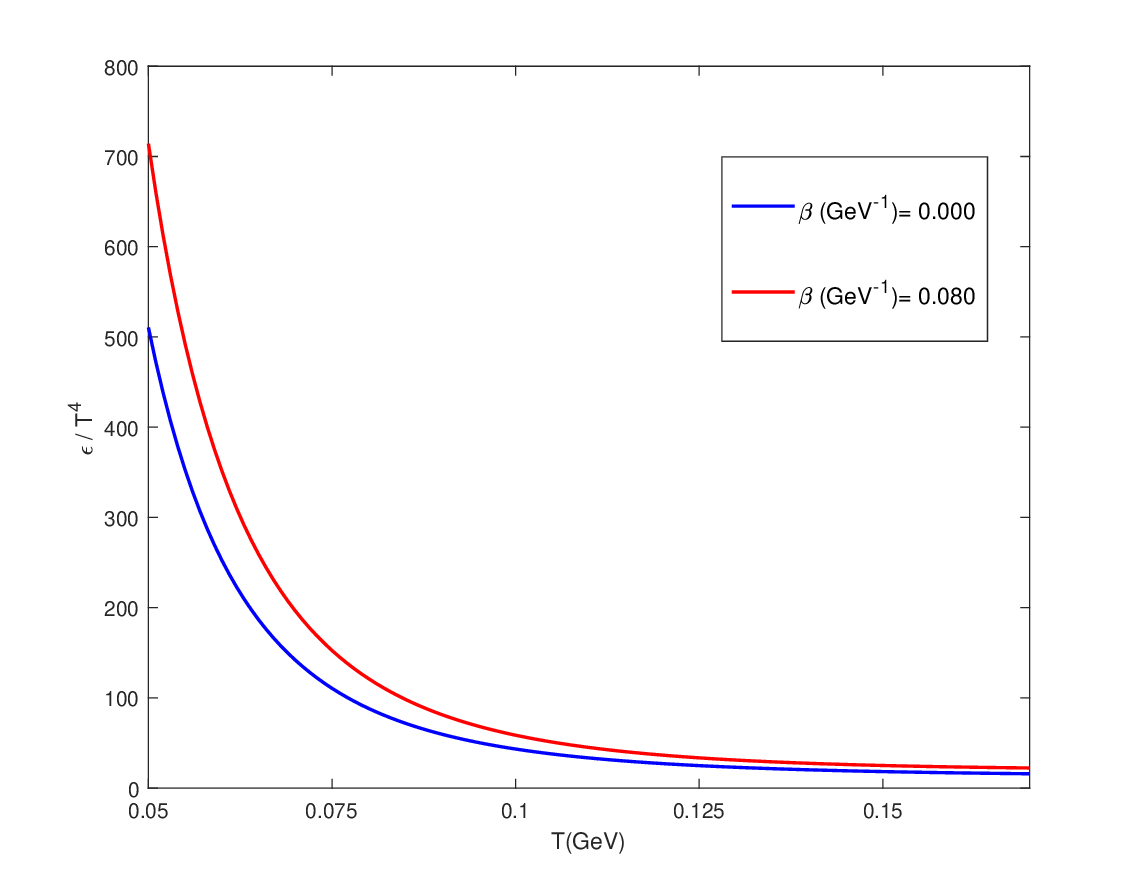}\hfill
	\caption{The energy density normalized to $T^4$ is given in dependence on T. }
	\label{epsioln}
\end{figure}
From the previously derived equations, a series of fundamental observations can be established:
\begin{itemize}
    \item 
The bag pressure $B$ (Eq. \ref{B}) is a positive quantity, leading to the pressure expression vanishing at a specific temperature $T_0$, which is given by:
    
  %  The bag pressor $B$ (Eq.\ref{B}) is a positive  quantity, which leads to the expression of the pressor vanishes for a temperature $T_0$ which is expressed by:

\begin{align}
   T_0 &= \frac{1}{2\beta^{2}\left(32g_g+31g_q\right)\frac{\pi ^4}{70}}
 \Bigg(-\beta\left(24g_g+\frac{45g_q}{2}\right)\frac{\zeta 
(5)}{\pi ^{2}}+\label{eq29}\\
&\sqrt{\left(\beta\left(24g_g+\frac{45g_q}{2}\right)\frac{\zeta 
(5)}{\pi ^{2}}\right)^2-4\beta^{2}\left(32g_g+31g_q\right)\frac{\pi ^{4}}{70}\left(\frac{\sigma_{SB}}{3}-B\right)}\Bigg)\nonumber
\end{align}
So for a positive pressor $P>0$, we need to verify that $T>T_0$.
\item 
According to Fig.(\ref{P}), which presents the pressor variation as a temperature function. %The degrees of freedom in the QGP state assuming $n_f = 2$ are  $g_{Q} =24$, $g_{g} = 16$ \big( \textcolor{red}{therefore $g_{QGP}=37$, since $g_{QGP}=g_{g}+\frac{7}{8}g_{q}$\big) and $g_{\pi} = 3$, \cite{yagi2005quark}}. 
We took $g_{QGP}=37$, The value $g_{QGP}=37$ arises from the effective degrees of freedom in the quark-gluon plasma (QGP) for $n_f=2$ light quark flavors and gluons. Quarks, being fermions, contribute $g_{Q} =24$ degrees of freedom, scaled by $\frac{7}{8}$  due to Fermi-Dirac statistics, yielding an effective contribution of $\frac{7}{8}g_{Q}=21$. Gluons, as bosons, contribute directly with$g_{g} = 16$. The total is 
$g_{QGP}=37$ \cite{yagi2005quark}. The factor $\frac{7}{8}$ appears due to the statistical mechanics of fermions. Fermions contribute differently to thermodynamic quantities like energy density and pressure compared to bosons. Specifically, this factor reflects the suppression of fermionic states due to the Pauli exclusion principle.

The figure demonstrates that the pressure expression reaches a maximum value $P_{max}/T^4=\sigma_{SB}$ for $T\rightarrow \infty$. Furthermore, it can be observed that the correction resulting from the LQGUP leads to an increase in pressure with rising temperature, such that $P\rightarrow \infty$ for $T \rightarrow \infty$. The fact that the presence of LQGUP leads to a violation of the well-known result, where the $P/T^{4}$ exceeds the Stefan-Boltzmann limit, at $T \rightarrow \infty$ may suggest that the effects of quantum gravity, as modeled by LQGUP, alter the nature of radiation and matter, in other words, the nature of radiation and matter no longer follows classical thermodynamics, due to QG. This implies that a new type of physics law could govern the behavior of matter and radiation at extreme conditions in the regime where quantum gravity becomes significant. However, experimental observation, if achievable, remains the definitive criterion for confirming this result.

%On the other hand, it can be noted that the correction resulting from the LQGUP led to the pressure increasing with increasing temperature so that ($P\rightarrow \infty$ for $T \rightarrow \infty$).  
\item 
Fig.( \ref{epsioln}) presents the energy density variation as a temperature function. The degrees of freedom in the QGP state assuming $n_f = 2$ are $g_{QGP} = 37$, $g_{Q} =24$, $g_{g} = 16$ and $g_{\pi} = 3$, \cite{yagi2005quark}.

 \end{itemize}

%\begin{equation}
%B=\left(g_{QGP}-g_{\pi} \right)\frac{\pi^2}{90}T_c^4.
%\end{equation}

%The bag model (BM) equation of state (EoS) has served as a foundational tool for describing the Quark-Gluon Plasma (QGP) for several decades. In its simplest form, assuming non-interacting, massless constituents and vanishing conserved charges, the BM EoS (prior to incorporating Generalized Uncertainty Principle (GUP) corrections) can be expressed as:
%$$ 
%\epsilon (T)=\sigma_{SB}T^4+B, \quad \quad P(T)=\frac{\sigma_{SB}}{3}T^4-B,
%$$
%where the energy density $\epsilon$ and the pressure $P$ have a simple dependence on temperature $T$ modified by adding a positive constant B (“vacuum pressure”). 

%For bosons (gluon) the density energy gives the well-known Stefan–Boltzmann result: 
%$$\epsilon_B=\frac{E_B}{3V}=\frac{g_B\pi^2}{30}T^4=3P_B,$$
%and for fermions a relative reduction factor, which is $\frac{7}{8}$ appears in the density energy:
%$$
%\epsilon_F=\frac{7}%{8}\frac{g_B\pi^2}{30}T^4=3P_F.
%$$

 Before we end this section, It is important to note here that in the region where the chemical potential is non-zero ($\mu\neq 0$), integrals of the form will appear in the expression of $\ln Z_{B}$ and $\ln Z_{F}$:\par 
$-\displaystyle \int_{0}^{+\infty} \ln(1-\exp(-x)\exp(\mu/T))x^{n}\sim \Li_{n+2}\left(\exp(\mu/T)\right)$, where $Li_{k}(x) $ are polylogarithm functions and since $\Li_{k}(x)$ are positive and increasing with $x$,   so, $\ln Z_{B}$ and $\ln Z_{F}$ are growing due to $\mu$, as a result,  one can expect that $\mu$ increases potential and energy density. Consequently, the impact of LQGUP will increase.

\section{Entropy and the speed of sound}
 In this section, we will derive the GUP corrections to the entropy density and the speed of the sound of an ideal QGP. First, we define the speed of sound in terms of the equation of state as follows \cite{yagi2005quark}:

\begin{equation}
 c_s^{2}=\frac{\partial P}{\partial \epsilon}=\frac{\partial \ln T}{\partial\ln s}  
\end{equation}
which can be rewritten as
\begin{equation}
 c_s^{2}=s\frac{\partial }{\partial s}\ln T= \frac{s}{T}\frac{\partial T}{\partial s}.
\end{equation}
Therefore, the speed of sound takes the form:
\begin{equation}
 c_s^{2}=\frac{s}{T} \left(\frac{\partial s }{\partial T} \right)^{-1} .
\label{eq32}
\end{equation}

In order to calculate the speed of sound using Eq. (\ref{eq32}), it is first necessary to determine the entropy density. For an ideal Quark-Gluon Plasma (QGP), the entropy density can be computed within the framework of the grand canonical ensemble. In this context, the entropy density is related to the grand potential, $\Omega$, through the following expression \cite{yagi2005quark}:
\begin{equation}
s=-\frac{1}{V}\frac{\partial \Omega}{\partial T}
\label{entr}
\end{equation}

Replacing  Eq. (\ref{eqq223}) in Eq. (\ref{entr}), The LQGUP modified entropy density of the system, will be:
\begin{equation}
s=\frac{2\pi^{2}}{45} \left(g_g+\frac{7}{8}g_q\right)T^{3}+\frac{ 15\zeta \left(
5\right)}{\pi ^{2}}\beta\left(8g_g+\frac{15g_q}{2}\right)T^{4}+\frac{3\pi ^{4}}{35}\beta^{2}\bigg(32g_g+31g_q\bigg)T^{5}.
\label{eqq19}
\end{equation}
In the limit $\beta \rightarrow 0$ (where $\beta$ is the GUP constant), the entropy density scales as $s \simeq T^3$, which recovers the Stefan–Boltzmann relation as expected in the absence of a GUP correction. This is readily evident from Eq. (\ref{eqq19}), which simplifies to:
\begin{align}
\left(\frac{\partial s}{\partial T}\right) &=\frac{6\pi^{2}}{45} \left(g_g+\frac{7}{8}g_q\right)T^{2}+\frac{ 60\zeta \left(
5\right)}{\pi ^{2}}\beta\left(8g_g+\frac{15g_q}{2}\right)T^{3}+\frac{15\pi ^{4}}{35}\beta^{2}\bigg(32g_g+31g_q\bigg)T^{4},
\\
\frac{s}{T}&=\frac{2\pi^{2}}{45} \left(g_g+\frac{7}{8}g_q\right)T^{2}+\frac{ 15\zeta \left(
5\right)}{\pi ^{2}}\beta\left(8g_g+\frac{15g_q}{2}\right)T^{3}+\frac{3\pi ^{4}}{35}\beta^{2}\bigg(32g_g+31g_q\bigg)T^{4},
\end{align}
and after substituting these equations in Eq. (\ref{eq32}), we obtain the LQGUP-modified speed of sound:
\begin{equation}
c_s^2=\frac{\left[\frac{2\pi^{2}}{45} \left(g_g+\frac{7}{8}g_q\right)\right] T^{2}+\left[\frac{ 15\zeta \left(
5\right)}{\pi ^{2}}\beta\left(8g_g+\frac{15g_q}{2}\right)\right]T^{3}+\left[\frac{3\pi ^{4}}{35}\beta^{2}\bigg(32g_g+31g_q\bigg)\right]T^{4}}{3\left[\frac{2\pi^{2}}{45} \left(g_g+\frac{7}{8}g_q\right)\right]T^{2}+4\left[\frac{ 15\zeta \left(
5\right)}{\pi ^{2}}\beta\left(8g_g+\frac{15g_q}{2}\right)\right]T^{3}+5\left[\frac{3\pi ^{4}}{35}\beta^{2}\bigg(32g_g+31g_q\bigg)\right]T^{4}}.
\label{eq37.}
\end{equation}
%\begin{figure}[h!]
%	\includegraphics[width=1\linewidth, height=0.48\textheight]{s.eps}\hfill
%	\caption{The entropy density is given in dependence on T. }
%	\label{s}
%\end{figure}
It is worth noting that there is a set of keynotes:%. Firstly, 
\begin{itemize}
    \item 
As the GUP constant $\beta$ approaches zero $(\beta\rightarrow 0)$, the speed of sound $(c^2_s) $ tends towards $\frac{1}{3}$. This recovers the expected value for the speed of sound in an ideal gas composed of massless particles, which aligns with the Stefan–Boltzmann limit.
\item 
 In the limit of infinite temperature $(T\rightarrow \infty)$, the speed of sound becomes $c_s^2 \rightarrow \frac{1}{5}$, which is different compared with the linear GUP modified speed of sound calculated in ref.\cite{demir2018effect}, 
 where $c_s^2\rightarrow\frac{1}{4}$ when $T\rightarrow \infty$. This suggests that the Linear or linear-quadratic GUP at high temperatures introduces a characteristic scale that disrupts the anticipated conformal invariance \cite{demir2018effect}. This symmetry breaking due to GUP, may mean that the AdS/CFT conjecture cannot be used to study QGP in the context of GUP.  
\end{itemize}
We expand the Eq.(\ref{eq37.}) in the second order of $\beta$ the limit $\beta^3 T^3<<1$, to obtain:
\begin{align}
c_s^2&\approx\frac{1}{3}-\frac{75\zeta(5)}{2\pi^4}\left(\frac{\frac{15}{2}g_q+8g_g}{\frac{7}{8}g_q+g_g}\right)\beta T+\frac{1}{27\left[\frac{2\pi^{2}}{45} \left(g_g+\frac{7}{8}g_q\right)\right]^2 } \nonumber\\
&\left(4\left[\frac{ 15\zeta \left(
5\right)}{\pi ^{2}}\left(8g_g+\frac{15g_q}{2}\right)\right]^2-6\left[\frac{2\pi^{2}}{45} \left(g_g+\frac{7}{8}g_q\right)\right]\left[\frac{3\pi ^{4}}{35}\bigg(32g_g+31g_q\bigg)\right]  \right) \beta^{2} T^2+\mathcal{O}(\beta^3 T^3)
\label{eq38}
\end{align}
Finally, an ideal QGP comprised of non-interacting gluons and three flavors of
massless quarks, the effective number of degrees of freedom for quarks and gluons are given as
$g_q = 36$ and $g_g = 16$, respectively. Therefore, the GUP-modified speed of sound, i.e., Eq. (\ref{eq38}),
for an ideal QGP with non-interacting gluons and three flavors of massless quarks reads:
\begin{align}
c_s^2&\approx\frac{1}{3}-\frac{5970\zeta(5)}{19\pi^4}\beta T+\frac{3}{361\pi^4}\left(\frac{900}{\pi^4}(398)^2\zeta(5)^2-\frac{36\pi^6}{1575}\frac{380}{8}1628    \right)\beta^2T^2\\
&\approx\frac{1}{3}-\frac{5970\zeta(5)}{19\pi^4}\beta T+\frac{3}{361}\left(\frac{900}{\pi^8}(398)^2\zeta(5)^2-\frac{61864}{35} \pi^2   \right)\beta^2T^2
\end{align}
Note that the correction to the Stefan–Boltzmann limit of the speed of sound squared is negative.

%\begin{gather}
%S=-\frac{\partial \Omega}{\partial T}\nonumber\\
%=\left(g_g+\frac{7}{8}g_q\right)\frac{2\pi^{2}V}{45} T^{3}+\beta\left(24g_q+\frac{45g_g}{2}\right)\frac{ 5VT^{4}\zeta \left(
%5\right)}{\pi ^{2}}+\beta^{2}\bigg(32g_q+31g_g\bigg)\frac{3V\pi ^{4}T^{5}}{35}.
%\end{gather}

\section{ Impact of speed of sound on bulk viscosity}
In the previous section, we determined the modified speed of sound, making it intriguing to examine how the GUP correction affects the bulk-to-shear viscosity ratio, $\frac{\zeta}{\eta}$. Buchel introduced a proposed bound on bulk viscosity in strongly coupled gauge plasmas in Ref. \cite{buchel2008bulk}:
%Since we calculated the modified speed of sound in the previous unit, it will be interesting to know the effects that the GUP correction has on the ratio of bulk viscosity to shear viscosity, i.e., $\frac{\zeta}{\eta}$. In Ref. \cite{buchel2008bulk} a bulk viscosity bound in strongly coupled gauge plasmas was proposed by Buchel:
\begin{equation}
 \frac{\zeta}{\eta} \geq 2\left(\frac{1}{3}-c_s^2\right)
 \label{soun}
\end{equation}
Note that this bound has a different dependence on the speed of sound from the relation for the
shear and bulk viscosity coefficients derived by Weinberg:
\begin{equation}
 \frac{\zeta}{\eta}=15\left(\frac{1}{3}-c_s^2\right)^2
 \label{eq422}
\end{equation}
By substituting the GUP-corrected speed of sound (see Eq. (\ref{eq38})),
the bulk viscosity bound, namely Eq. (\ref{soun}), now reads:
\begin{align}
 &\frac{\zeta}{\eta}\geq \frac{75\zeta(5)}{\pi^4}\left(\frac{\frac{15}{2}g_q+8g_g}{\frac{7}{8}g_q+g_g}\right)\beta T+\nonumber \\
 &\frac{2}{27\left[\frac{2\pi^{2}}{45} \left(g_g+\frac{7}{8}g_q\right)\right]^2 }\left[\frac{36\pi^{6}}{1575} \left(g_g+\frac{7}{8}g_q\right)\bigg(32g_g+31g_q\bigg) -\frac{ 900\zeta^2 \left(
5\right)}{\pi ^{4}} \left(8g_g+\frac{15g_q}{2}\right)^2 \right] \beta^{2} T^2
\label{sogup}
\end{align}
Based on Eq. (\ref{sogup}), and for any non-zero temperature $T$, the presence of the GUP deformation introduces a non-zero minimum value for the bulk viscosity, $\zeta$. Without the GUP correction $(\beta \rightarrow 0)$, the bound simplifies to $\zeta/\eta \geq 0$. This aligns with the expectation of $\zeta = 0$ for conformally invariant systems, such as those composed of non-interacting massless particles \cite{demir2018effect}. However, the emergence of a non-zero minimum for $\zeta$ at non-zero temperatures when the GUP correction is included suggests that the GUP breaks the a priori conformal invariance of the system by introducing a new characteristic scale.

Replacing Eq. (\ref{eq38}) in Eq. (\ref{eq422}) we get the bulk viscosity for an ideal QGP.

\begin{align}
\frac{\zeta}{\eta}\approx 15 \Bigg[\frac{75\zeta(5)}{2\pi^4}\left(\frac{\frac{15}{2}g_q+8g_g}{\frac{7}{8}g_q+g_g}\right)\beta T+
 75 \left[\frac{12\pi^{2}}{1575} \bigg(\frac{32g_g+31g_q}{g_g+\frac{7}{8}g_q}\bigg) -\frac{ 450\zeta^2 \left(
5\right)}{\pi ^{8}} \left(\frac{8g_g+15g_q/2}{g_g+\frac{7}{8}g_q}\right)^2 \right] \beta^{2} T^2\Bigg]^2
\end{align}
If we took the expansion of Eq. (\ref{sogup}) to the second order of $\beta$, we found the same result as the one published in (Eq. (39) \cite{demir2018effect}).

Evaluating Eq. (\ref{eq45}) for an ideal QGP of noninteracting gluons and three flavors of massless
quarks ($g_g = 16$ and $g_q = 36$) yields:
\begin{equation}
\frac{\zeta}{\eta}\approx15\left[\frac{5970\zeta(5)}{19\pi^4}\beta T+\frac{3}{361}\left(\frac{61864}{35} \pi^2   -\frac{900}{\pi^8}(398)^2\zeta(5)^2\right)\beta^2T^2
\right]^2
\label{eq45}
\end{equation}
The results of the final calculations provide us with the following insights:
\begin{itemize}
    \item 
In the absence of the GUP deformation, $(\beta\rightarrow 0)$ the speed of sound is $c_s=\frac{1}{3}$, so the bulk viscosity coefficient approaches zero, i.e., $\zeta \rightarrow 0$, This aligns with the behavior observed in an ideal gas composed of massless particles.
\item 
Based on Eq. (\ref{eq45}) the GUP could potentially introduce a new characteristic scale into the system, thereby violating conformal invariance. This effect might be particularly relevant for bulk mediums with short mean free paths dominated by radiative quanta.
\item 
In the limit $T\rightarrow \infty$, the speed of sound becomes $c_s^2\rightarrow \frac{1}{5}$, accordingly modifying the bulk viscosity bound i.e., Eq. (\ref{soun}), to:
\begin{equation}
\frac{\zeta}{\eta} \geq \frac{4}{15}
\label{vis}
\end{equation}
Whereas, the lower limit of Eq. (\ref{vis}), gives the same value predicted using Weinberg’s
formula, since as $c_s^2\rightarrow \frac{1}{5}$, Eq.(\ref{eq422}) implies $\frac{\zeta}{\eta}\rightarrow \frac{4}{15} $, therefore in the limit $T\rightarrow \infty$
\begin{equation}
\left(\frac{\zeta}{\eta}\right)_{Weinberg}=\left(\frac{\zeta}{\eta}\right)_{Buchel, min}\label{eqq}
\end{equation}
\end{itemize}
%In contrast to the result obtained in (Ref. \cite{demir2018effect}), in which there study was based on the Linear GUP deformation, the Buchel lower limit of the bulk viscosity coefficient expression was found to be larger than the one calculated using Weinberg formula Eq. (\ref{eq422}). Based on our calculation using GUP Liear-quadratic GUP deformation. The obtained result Eq. (\ref{eqq}), we found that: at the limit $T\rightarrow \infty$ the two expressions leads to the same value. 
Our findings contrast with those presented in Ref. \cite{demir2018effect}, where a Linear GUP deformation yielded a lower limit for the bulk viscosity coefficient exceeding the value obtained using the Weinberg formula (Eq. (\ref{eq422})). Utilizing the Linear-quadratic GUP deformation in our analysis (Eq. (\ref{eqq})), we observe that the two expressions converge to the same value in the limit $T\rightarrow \infty$.
\section{The  time Evolution of the Temperature and Energy Density of Early Universe}
Observations indicate that the universe is approximately homogeneous and isotropic when averaged over scales much larger than the size of galaxies and galaxy clusters. Therefore, one may take the Roberston-Walker  (RW) metric as a good discretion of the global structure of the universe \cite{ornik1987expansion,yagi2005quark}. Consequently, the energy density, pressure, and temperature of the universe are functions of time $t$ only; $\varepsilon(t), P(t), T(t)$. Under these considerations, the Friedmann equation is written as
\begin{gather}
H^{2}=\left(\frac{\dot{a}}{a}\right)^{2}=\frac{8\pi G}{3}\epsilon-\frac{K}{a^{2}}, \label{Hubble}
\end{gather}
$H$ is called "Hubble parameter", $a(t)$ is the scale factor and $\dot{a}(t)$ is the derivative of scale factor with time.  $G$ is the gravitational constant, $\epsilon$ is the energy density of the universe, and $K$ is the sing of the spatial curvate, where $K=0, -1, \text{and} +1$ indicates the flat space, closed space with positive curvature, and open space with negative curvature, respectively. 
\begin{gather}
\frac{d\epsilon}{da}=-\frac{3}{a}\left(\epsilon+P\right), \label{scale}
\end{gather}
where $P$ is the pressure. For the QGP phase, we have (see the Eqs. (\ref{pressure}) and  (\ref{density})) 

\begin{gather}
P= A-B, \\
\epsilon= 3A+B
\end{gather}
where $A$ is written as 
\begin{gather}
A=\left(g_g+\frac{7}{8}g_q\right)\frac{\pi^{2}}{90} T^{4}+\beta\left(24g_g+\frac{45g_q}{2}\right)\frac{ T^{5}\zeta \left(
5\right)}{\pi ^{2}}+\beta^{2}\bigg(32g_g+31g_q\bigg)\frac{\pi ^{4}T^{6}}{70}
\end{gather}
so, $\epsilon=3P+4B$,  as a consequence the Eq. (\ref{scale}) becomes
\begin{gather}
\frac{d\epsilon}{da}=-\frac{4}{a}\left(\epsilon-B\right).
\end{gather}
this simple one-order differential equation can be solved by putting  the following special solution
\begin{gather}
\epsilon=\frac{1}{a^{4}}+B. 
\end{gather}
on the other hand, we can get the $a$ in terms of energy density $\epsilon$;
$a=1/\left(\epsilon-B\right)^{1/4}$, and by replacing this relation in Eq. (\ref{Hubble}), the evolution of energy density with time is written as
\begin{gather}
-\frac{\dot{\epsilon}}{4(\epsilon-B)}=\sqrt{\frac{8\pi G}{3}\epsilon-K \epsilon^{1/2}}.
\end{gather}
We can neglect $K$ in the early universe, and as we mentioned in the introduction, the QGP formed and it is important only in the early universe, so for our sake, we take $K=0$ \cite{elmashad2021effect, ornik1987expansion}. Therefore we conclude the following differential equation of energy density evolution with time
\begin{gather}
-\frac{\dot{\epsilon}}{4(\epsilon-B)}=\sqrt{\frac{8\pi G}{3}\epsilon}.
\end{gather}

%\begin{figure}[H]
	%\includegraphics[width=1\linewidth, height=0.48\textheight]{Delta.eps}\hfill
	%\caption{The variation of the LGGUP-term of the energy density of an ideal QGP versus time. }
	%\label{Delta}
%\end{figure}

The solution of this simple differential equation is written as
\begin{gather}
\epsilon(t)=B\tanh^{2}\left(\sqrt{\dfrac{32\pi B G}{3}} t+\arctanh\left(\sqrt{\frac{\epsilon(0)}{B}}\right) \right).\label{ep}
\end{gather}
  $\tanh(x)$ is the hyperbolic tangent function, and $\arctanh(x)$ is the inverse of  the hyperbolic tangent function.  If we consider that the energy density is divergent at $t=0$, i.e. $\epsilon(0) \rightarrow +\infty$ then one can simplify the solution (\ref{ep}) to become \cite{rafelski2013connecting, florkowski2011realistic}, 
  \begin{gather}
    \epsilon(t)=B\coth^{2}\left(\sqrt{\dfrac{32\pi B G}{3}} t \right).\label{efp}
  \end{gather}
  $\coth(x)$ is the inverse function of $\tanh(x)$. 
  The effect of quantum gravity (GUP) on the energy density evolution lies in the change in the value of bag pressure $B$, where according to the expression (\ref{B}), $B$ is affected by the GUP parameter, $\beta$. Finding the analytical expression of the temperature $T(t)$, within the context of GUP is very complicated, where according to Eq. (\ref{density}), one must solve the sixth order equation, so, we will plot the time evolution of $T$ directly.  It is important to notice, that the evolution of energy density (\ref{efp}) is only valid in the range $0< t < t_{I}$, where $T(t_{I})=T_{c}$, for example, for $T_{c}=170 \text{MeV}$, $t_{I}=18\mu s$ \cite{yagi2005quark}, and of course, since we focus on QGP phase, the temperature must be bigger than the critical temperature $T_{c}$. Consequently, we cannot plot the energy density evolution, until, we get $t_{I}$ for $\beta=0\hspace{0.1cm}\text{GeV}^{-1}$ and $\beta\neq 0$ for our case.  \par
\begin{figure}[h!]
	\includegraphics[width=1\linewidth, height=0.48\textheight]{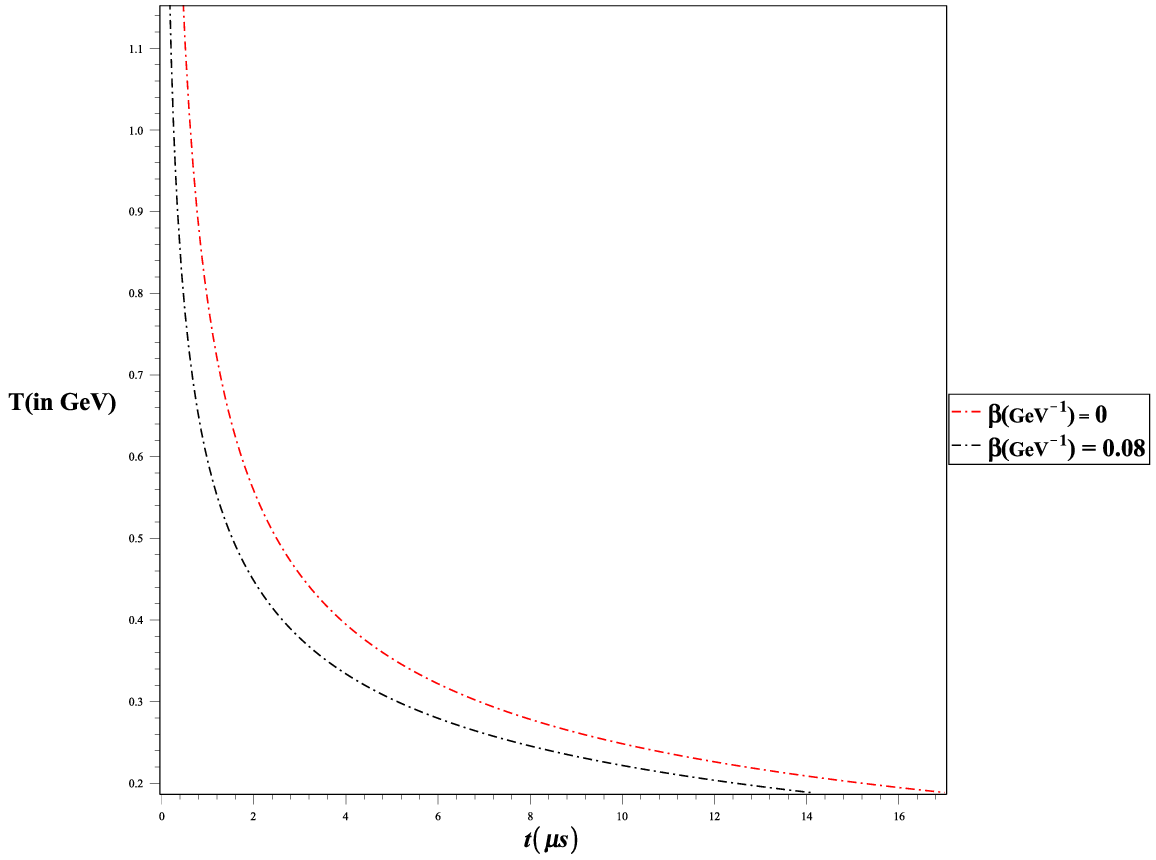}\hfill
	\caption{The temperature versus time with the GUP-effect (dotted dashed black),
without GUP-effect (dotted dashed red). The parameters are $B\left(\beta=0\hspace{0.1cm} \text{GeV}^{-1}, T_{\text{c}}=0.170\hspace{0.1cm} \text{GeV}\right) = 0.0031\hspace{0.1cm} \text{GeV}^{-4}$, $B\left(\beta=0.08\hspace{0.1cm} \text{GeV}^{-1}, T_{\text{c}}=0.170 \hspace{0.1cm}\text{GeV}\right)=0.0044\hspace{0.1cm}\text{GeV}^{-4}$}
	\label{Delta2}
\end{figure}

 
  %The  difference in energy density evolution  $\Delta \epsilon (t)=\epsilon_{\beta=0.005}(t)-\epsilon_{\beta=0}(t)$ is plotted in fig. (\ref{Delta}). It is observed that the quantum gravity or the GUP with the form (\ref{gupp}) increases the energy density, and the difference between the two cases (with and without GUP) increases with time.\par
  By replacing the energy density evolution expression (\ref{efp}) in the expression (\ref{density}), one can deduce the evolution of temperature. In fig. (\ref{Delta2}), we presented $T(t)$ in the absence and the presence of GUP.

One can see that the GUP decreases the temperature of the QGP phase. Consequently, the time $t_{I}$ where the temperature of the QGP reaches the critical temperature $T_{c}=0.170 \hspace{0.1cm} \text{GeV}$ is reduced under the impact of quantum gravity, where,  for $\beta=0 \hspace{0.1cm}\text{GeV}^{-1}$, the time that the system takes to transform from QGP phase to hadronic phase is  $t_{I}=20.62\hspace{0.1cm}\mu s$, while for $\beta=0.08 \hspace{0.1cm}\text{GeV}^{-1}$, $t_{I}=17.39\hspace{0.1cm}\mu s$.\par 
Now, since we got the values of $t_{I}$, one can present $\epsilon(t)$, where to compare between the energy density evolution for $\beta=0$ and $\beta=0.08\hspace{0.1cm}\text{GeV}^{-1}$, the interval of time must be in the range $[0, 17.36 \mu s]$. %The presentations of $\epsilon(t)$ for $\beta=0\hspace{0.1cm}\text{GeV}^{-1}$ and $\beta=\hspace{0.1cm} 0.08\text{GeV}^{-1}$  are presented in the fig. \ref{energy}
%\begin{figure}[H]
	%\includegraphics[width=1\linewidth, height=0.48\textheight]{densityyy.eps}\hfill
	%\caption{The variation of the LGGUP-term of the energy density of an ideal QGP versus time, for $\beta=0\hspace{0.1cm}\text{GeV}^{-1}$ and $\beta=0.08\hspace{0.1cm}\text{GeV}^{-1}$. The parameters are $B\left(\beta=0\hspace{0.1cm} \text{GeV}^{-1}, T_{\text{c}}=0.170\hspace{0.1cm} \text{GeV}\right) = 0.0031\hspace{0.1cm} \text{GeV}^{-4}$, $B\left(\beta=0.08\hspace{0.1cm} \text{GeV}^{-1}, T_{\text{c}}=0.170 \hspace{0.1cm}\text{GeV}\right)=0.0044\hspace{0.1cm}\text{GeV}^{-4}$ }
	\label{energy}
%\end{figure}
 We will illustrate the difference $\Delta (\epsilon)=\epsilon(\beta=0.08\hspace{0.1cm}\text{GeV}^{-1})-\epsilon(\beta=0\hspace{0.1cm}\text{GeV}^{-1})$ to understand how quantum gravity affects the evolution of the energy density for the QGP phase.
\begin{figure}[H]
	\includegraphics[width=1\linewidth, height=0.48\textheight]{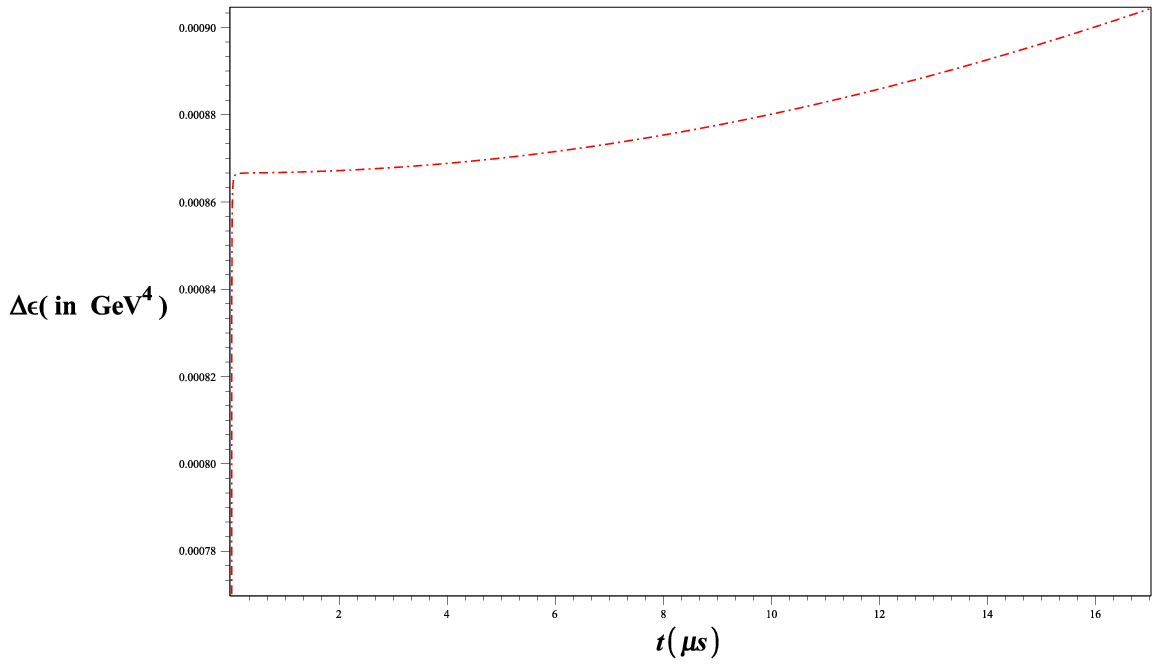}\hfill
	\caption{The variation of the LGGUP-term of the energy density of an ideal QGP versus time, for $\beta=0\hspace{0.1cm}\text{GeV}^{-1}$ and $\beta=0.08\hspace{0.1cm}\text{GeV}^{-1}$. The parameters are $B\left(\beta=0\hspace{0.1cm} \text{GeV}^{-1}, T_{\text{c}}=0.170\hspace{0.1cm} \text{GeV}\right) = 0.0031\hspace{0.1cm} \text{GeV}^{-4}$, $B\left(\beta=0.08\hspace{0.1cm} \text{GeV}^{-1}, T_{\text{c}}=0.170 \hspace{0.1cm}\text{GeV}\right)=0.0044\hspace{0.1cm}\text{GeV}^{-4}$.}
	\label{energyu}
\end{figure}
One can notice that $\Delta \epsilon(t)$ is positive which means that the LQGUP (or quantum gravity) increases the energy density of the QGP phase. The growth rate due to the effect of LQGUP is maximal in very early time $t\in [0, 1 \mu s]$. Moreover, since $\Delta \epsilon(t)$ grows with time, one can say that the influence of quantum gravity increases with time. \par 
In the introduction, we mentioned that high-energy collisions can produce the QGP, and in this section, we studied QGP in the early universe, Thus, it may be important to highlight the aspects of the similarities/differences \cite{yagi2005quark}:\par
According to Friedmann's solution to Einstein's gravitational equation, the universe is expanding. Tracing this expansion backward in time reveals that as the universe contracts, matter, and radiation become increasingly hot and dense, eventually reaching conditions suitable for the formation of a Quark-Gluon Plasma (QGP).\par
In contrast, high-energy collisions achieve similar high temperatures and densities through heavy-ion accelerators. These accelerators propel nuclei to energies of several tens of GeV, causing them to collide and interact intensely, remaining nearby.

\section{Conclusion}
In this manuscript, we have conducted a study on calculating the thermodynamic properties of a QGP, modeled initially as an ideal
chemically equilibrated gas of quarks and gluons, including the effect of
confining vacuum structure. In the study of the quark-and-gluon gas,
our task was considerably simplified by the observation that the gluons and quarks are to all intent massless particles, at least on the
scale of energies available in the hot plasma $T=200 $ MeV.  In addition,  we found that the GUP decreases the evolution of QGP temperature, which means the duration required for the system to transition from the QGP phase to the hadronic phase is reduced under the influence of quantum gravity. On the other hand, the LQGUP increases the evolution of energy density, where this impact of LQGUP grows with time.

\bibliographystyle{unsrt} 
\bibliography{ref}

\end{document}